\documentclass[aps,prl,twocolumn,superscriptaddress,showpacs]{revtex4-1}
\usepackage{hyperref}
\usepackage{graphicx}
\usepackage{epsfig}
\usepackage{subfigure}
\usepackage{setspace}
\usepackage{amsmath}

\begin{document}

\title{Quantum dot nonlinearity through cavity-enhanced feedback with a charge memory}

\author{Morten P. Bakker}
\affiliation{Huygens-Kamerlingh Onnes Laboratory, Leiden University, P.O. Box 9504, 2300 RA Leiden, The Netherlands}
\author{Thomas Ruytenberg}
\affiliation{Huygens-Kamerlingh Onnes Laboratory, Leiden University, P.O. Box 9504, 2300 RA Leiden, The Netherlands}
\author{Wolfgang L\"{o}ffler}
\affiliation{Huygens-Kamerlingh Onnes Laboratory, Leiden University, P.O. Box 9504, 2300 RA Leiden, The Netherlands}
\author{Ajit Barve}
\affiliation{University of California Santa Barbara, Santa Barbara, California 93106, USA}
\author{Larry Coldren}
\affiliation{University of California Santa Barbara, Santa Barbara, California 93106, USA}
\author{Martin P. van Exter}
\affiliation{Huygens-Kamerlingh Onnes Laboratory, Leiden University, P.O. Box 9504, 2300 RA Leiden, The Netherlands}
\author{Dirk Bouwmeester}
\affiliation{Huygens-Kamerlingh Onnes Laboratory, Leiden University, P.O. Box 9504, 2300 RA Leiden, The Netherlands}
\affiliation{University of California Santa Barbara, Santa Barbara, California 93106, USA}

\date{\today}

\begin{abstract}
In an oxide apertured quantum dot (QD) micropillar cavity-QED system, we found strong QD hysteresis effects and lineshape modifications even at very low intensities corresponding to $<10^{-3}$ intracavity photons.
We attribute this to the excitation of charges by the intracavity field; charges that get trapped at the oxide aperture, where they screen the internal electric field and blueshift the QD transition.
This in turn strongly modulates light absorption by cavity QED effects, eventually leading to the observed hysteresis and lineshape modifications.
The cavity also enables us to observe the QD dynamics in real time, and all experimental data agrees well with a power-law charging model.
This effect can serve as a novel tuning mechanism for quantum dots.
\end{abstract}

\maketitle

Cavity quantum electrodynamics with quantum dots (QDs) coupled to microcavities enables various applications such as single-photon switches \cite{Chang2007,Englund2012,Bose2012,Loo2012,Volz2012}, generation of non-classical states of light \cite{Fushman2008,Majumdar2012,Munoz2014} and hybrid quantum information schemes \cite{Kimble2008,Bonato2010}.
However, QDs deviate from an ideal atom-like systems as they strongly interact with their environment, for example through nuclear spins \cite{Urbaszek2013, Chekhovich2013b} and via charge traps \cite{Houel2012,Kuhlmann2013}.
These interactions need to be understood and controlled in order to improve the QD coherence properties.
For this purpose cavities are very useful to probe the QD environment, through increased light-matter interaction.
\begin{figure}[b]
\centering
\centerline{\includegraphics[angle=0]{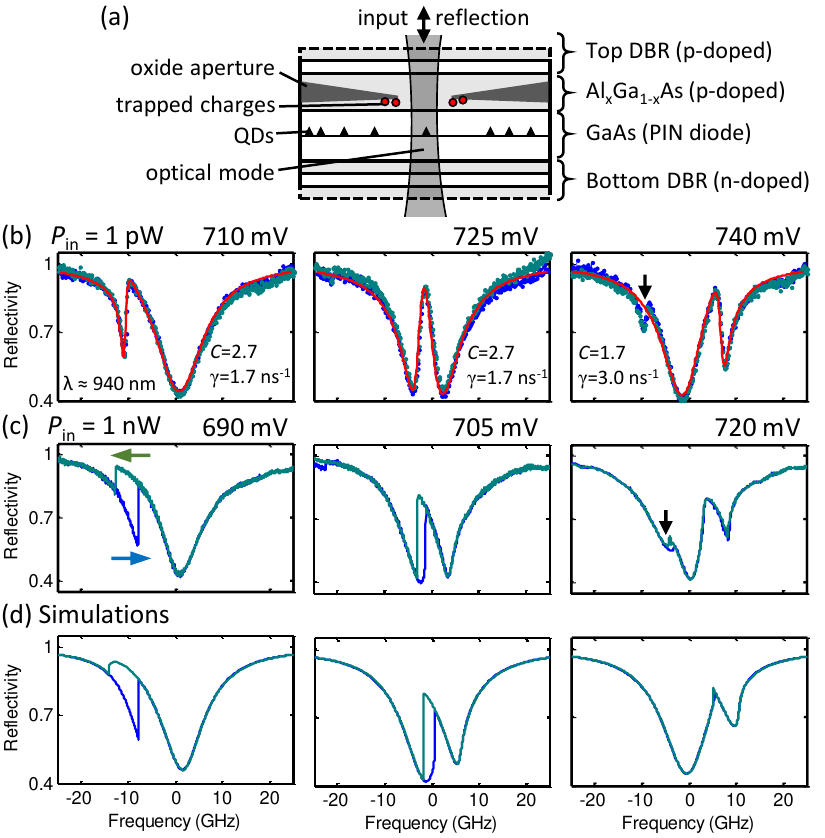}}
\caption{(a) Schematic of the sample structure with charges trapped at the oxide aperture. Figure is not to scale and only a couple DBR pairs are shown.
Resonant reflection spectroscopy scans recorded using laser intensities of (b) 1pW and (c) 1 nW for various applied bias voltages. Blue (green) curve: upward (downward) frequency scan. Red lines in (b) are fits using Eqn. \ref{Eq1} and reflectivity $R=|1-t|^2$. The QD cooperativity $C$ and dephasing rate $\gamma$ obtained from the fits are named in the figures. The black arrows denote a second QD in the same cavity. Scans were taken on $\sim$s timescale. (d) presents simulations that predict the scans in (c).}
\label{Fig1}
\end{figure}

In this Letter we investigate such a QD-cavity system.
For sufficiently low optical field intensity this system can be described by the QED of an effective 2-level system in a single-mode cavity.
For increasing intensities we report on bistable and strong nonlinear behavior.
The sample under study consists of InAs self-assembled QDs inside a PIN diode structure embedded in a micropillar.
This system combines QD charge and Stark shift control by applying a bias voltage with high-quality polarization-degenerate cavity modes \cite{Stoltz2005,Strauf2007,Rakher2009,Bonato2009,Bakker2014}.
The mode confinement in the transversal direction is achieved by an oxide aperture formed through a wet oxidation step.
The observed bistability and nonlinear behavior in the cavity QED system can be explained by attributing a second role to the oxide aperture, namely that of a charge memory.
Charges in this memory, created by resonant absorption, will cause a modification to the applied bias voltage which in turn shifts the QD frequency, and modify the amount of absorption.
In Figure~\ref{Fig1}~(a) the sample structure, with charges trapped at the oxide aperture, is schematically shown.

We consider one of the fine-split transitions of a charge neutral QD coupled to a polarization degenerate cavity mode in the intermediate coupling regime.
Figure \ref{Fig1} (b) shows reflection spectra, recorded at a sufficiently low incident intensity $P_{in} = 1$ pW such that no nonlinear effects occur.
Upward and downward frequency scans overlap perfectly and can be fitted by theory for a dipole inside an optical cavity, which we will discuss later in detail.
However, when a higher intensity of 1 nW is used, several strong deviations occur, see Fig. \ref{Fig1} (c).
First of all, a hysteresis feature appears when the QD is tuned at or below the cavity resonance.
Second, while at the high frequency side of the cavity resonance the hysteresis is much less, a line shape modification is still visible.
Finally, in order to obtain the same QD detuning compared to the low intensity scans, a lower bias voltage has to be applied.
After a thorough characterization of this effect, we present a model that explains all features as is shown in Fig. \ref{Fig1} (d).

\begin{figure}[b]
\centering
\centerline{\includegraphics[angle=0]{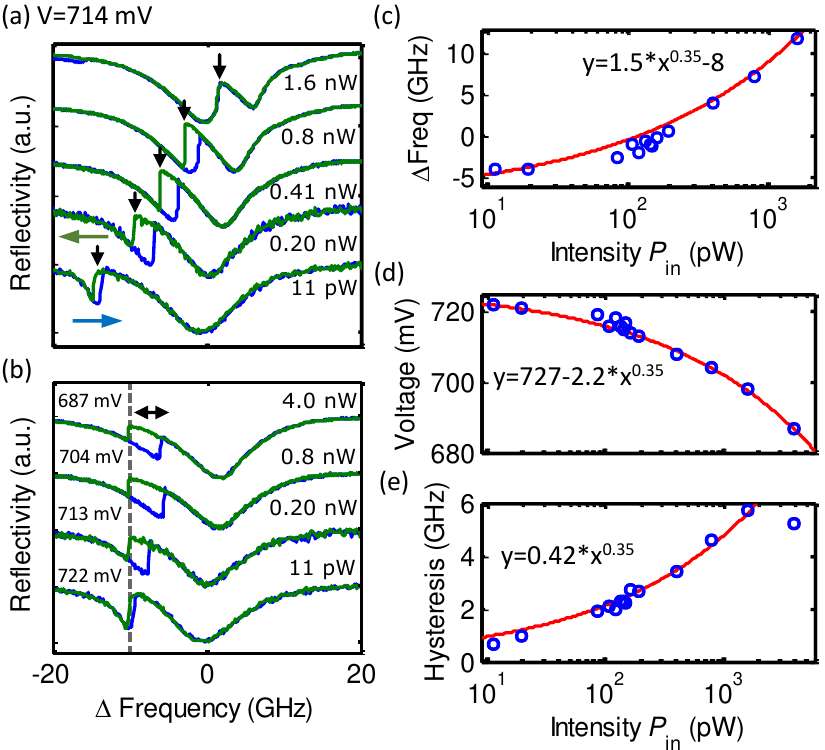}}
\caption{Characterization of the QD blueshift and nonlinearity. (a) Resonant reflection spectra for a fixed bias voltage and various intensities of the scanning laser. Blue (green): low to high (high to low) frequency scan. (b) Scans where the QD-cavity detuning is kept constant by varying the bias voltage for various laser intensities. (c) Relative QD shift (estimated from the vertical arrows in (a)), (d) applied bias voltage to keep the QD-cavity detuning constant (vertical dashed line in (b)), and (e) hysteresis width (horizontal arrow in (b)), all as function of the laser intensity. Red lines in (c,d,e) show empirical power-law fits. Vertical offsets have been added to the scans in (a,b).}
\label{Fig2}
\end{figure}

A first hint on the underlying dynamics is provided by investigating the power dependence in single-laser scans.
In Fig. \ref{Fig2} (a) we first keep the QD bias voltage constant and show scans for increasing laser intensity.
A QD blueshift occurs, as is displayed in Fig. \ref{Fig2} (c).
In a second set of measurements in Fig. \ref{Fig2} (b), we keep the QD-cavity detuning constant by changing the bias voltage as function of the laser intensity.
A lower bias voltage has to be applied for increasing laser intensity, shown in Fig. \ref{Fig2} (d).
Furthermore, the hysteresis width increases with intensity, see Fig. \ref{Fig2} (e), up to nearly 6 GHz, but then saturates and even decreases slightly when the intensity is above the QD saturation intensity ($\sim2.5$ nW).
All three observations in Figs. \ref{Fig2} (c-e) obey the same empirical power-law $\propto P_{in}^{\beta}$, with $\beta = 0.35$.
Already at an incident intensity $P_{in} = 11$ pW, corresponding to a maximum mean photon per cavity lifetime of $\langle \overline{n}\rangle\sim4\times10^{-4}$, a QD blueshift and line shape modification is clearly visible.
Here the mean intracavity photon number is found from $\langle\overline{n}\rangle=P_{out}/\kappa_m\hbar\omega$, where $\kappa_m \sim 11$ ns$^{-1}$ is the mirror loss rate, $\omega$ is the light angular frequency, and the maximum output intensity $P_{out} = |t|^2 P_{in}\sim1$ pW (see below).

\begin{figure}[h]
\centering
\centerline{\includegraphics[angle=0]{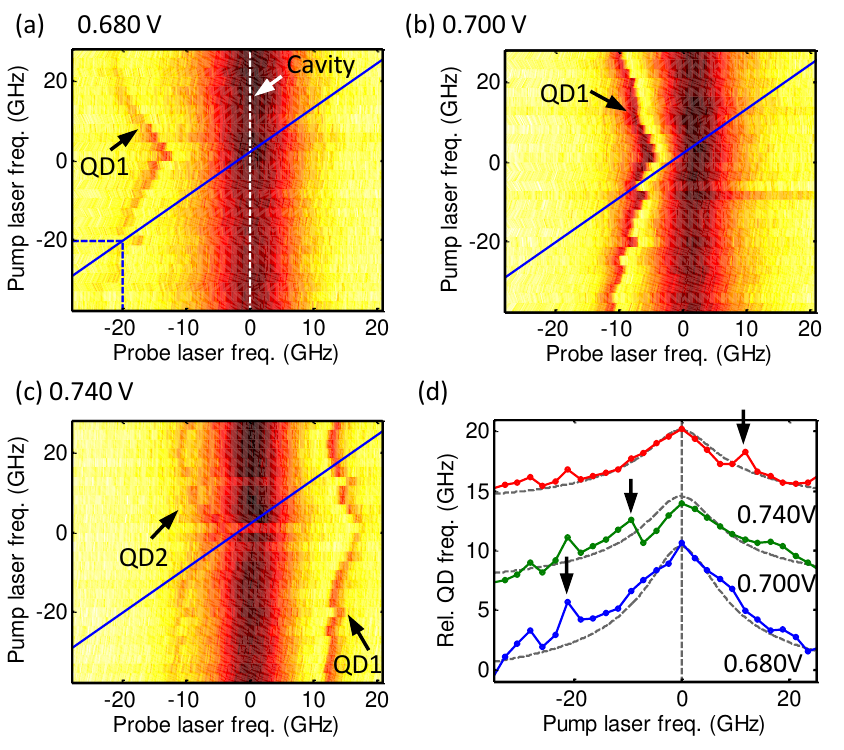}}
\caption{Two-laser scans with resonant lasers. (a-c) Reflectivity color maps as function of the weak (1 pW) probe laser frequency and the second high intensity (1 nW) pump laser frequency for various applied bias voltages. The blue line shows the pump laser frequency compared to the probe laser. The arrows indicate the QD position. (d) Relative QD frequency shift as function of the pump laser-cavity detuning. The arrows denote where the laser is resonant with the QD. Gray dashed lines: Lorentzian function convoluted with the power-law from Fig.~\ref{Fig2}~(c) that have been added as a reference. Vertical offsets are added to the curves.}
\label{Fig3}
\end{figure}

We now switch to a two-laser experiment (see Fig. \ref{Fig3}) in order to further investigate the phenomenon that the QD blueshifts with increasing laser intensity.
The QD transition is probed with a low intensity (1 pW) probe laser such that no line shape modification occurs.
We then add a second high intensity (1 nW) pump laser with orthogonal polarization, such that it can be filtered out with a crossed polarizer in the detection channel.
The pump laser is scanned in steps across the cavity resonance; for each step the QD-cavity spectrum is measured with the weak probe laser.
Figure \ref{Fig3} (a-c) presents these measurements for various bias voltages such that the average QD-cavity detuning is varied.
For every bias voltage the QD-cavity and pump laser-cavity detunings are determined, and the relative QD frequency as function of the pump laser-cavity detuning is shown in Fig. \ref{Fig3} (d).
The maximum QD blueshift (of about 8 GHz) occurs when the pump laser is resonant with the cavity mode, corresponding to the vertical dashed line.
An increased blueshift also occurs when the pump laser is close to the QD frequency, indicated by the arrows in Fig. \ref{Fig3} (d), at which point the intracavity field also increases due to cavity QED effects.
The gray dashed curves are Lorentzian lines convoluted with the $\propto P_{in}^{0.35}$ power-law from Fig.~\ref{Fig2}~(c) and correspond to the data nicely.
In conclusion, the data clearly shows that the QD resonance blueshifts when the intracavity field increases, and this effect is independent of the QD-laser or the QD-cavity detuning.

We now introduce a model that explains the dynamic line shape modifications.
We start from the transmission amplitude of a cavity with a coupled dipole \cite{auffeves2007,waks2006}:
\begin{equation}\label{Eq1}
t = \eta_{out}\frac{1}{1-i\Delta+\frac{2C}{1-i\Delta'}},
\end{equation}
where $\Delta = 2(\omega-\omega_c)/\kappa$ is the relative detuning between the laser ($\omega$) and cavity ($\omega_c$) angular frequencies, $C$ is the device cooperativity, $\Delta' = (\omega-\omega_{QD})/\gamma$ is the relative detuning between the laser and QD transition ($\omega_{QD}$), $\eta_{out}$ is the output coupling efficiency, $\kappa$ is the total intensity damping rate of the cavity and $\gamma$ is the QD dephasing rate.
We obtain close to perfect mode-matching, and therefore the total transmittivity through the cavity is given by $T = |t|^2$, and the total reflectivity is given by $R = |1-t|^2$.
In Fig.~\ref{Fig1}~(a) we show that the model fits the low intensity measurements very well.
The QD cooperativity and dephasing parameters $C$ and $\gamma$ are noted in the subfigure windows, and a cavity damping rate $\kappa \sim 77$ ns$^{-1}$ and an output coupling efficiency $\eta_{out} =2\kappa_m/\kappa\sim 0.3$ is found, corresponding to a cavity $Q$-factor of $Q\sim2.6\times10^4$.

As a next step, we introduce a QD frequency $f_{QD} = \omega_{QD}/2\pi$ that dynamically changes with the intracavity intensity, which is proportional to $|t|^2$:
\begin{equation}\label{Eq2}
f_{QD} = f_{0} + \alpha |t|^{2\beta},
\end{equation}
where $f_0$ equals the QD frequency in the limit of vanishing intracavity intensity.
Based on the empirical values determined in Fig.~\ref{Fig2}~(c), we use $\alpha=1.5\times(P_{in}/\eta_{out}^2)^\beta$ GHz, with $P_{in}$ in pW, and $\beta=0.35$.
Due to cavity QED effects, the intracavity field depends strongly on the QD frequency such that the cavity can change from being largely transparent to being largely reflective through only small changes in the QD frequency.
This interplay leads to the observed nonlinear behaviour.

Finally, we take into account that we are operating the QD--cavity system close to the saturation intensity, which slightly suppresses the QD features.
We take this into account by calculating the reflectivity $R'$ by taking the weighted sum of the reflectivities of a coupled ($R$) and uncoupled ($R_0$) cavity: $R' = xR + (1-x)R_0$, with $x=0.8$.
This is a strong simplification of a more rigorous approach based on the quantum master equation and calculation of the intracavity photon number \cite{Loo2012}, but is sufficient for our purpose.
For QD cooperativity $C$ and dephasing rate $\gamma$ we use the values obtained in the fits in Fig.~\ref{Fig1}~(a).
Figure~\ref{Fig1}~(c) shows that the predicted scans match the actual measurements very well.

As the underlying physical mechanism we hypothesize that charges are excited by the resonant laser and get trapped at the oxide aperture.
Reasonably high doping concentrations were used, up to $6\times10^{18}$ cm$^{-3}$ for the carbon p-doped layers and up to $5\times10^{18}$ cm$^{-3}$ for the silicon n-doped layers, and it is well-known that absorption takes place in these doped layers \cite{Sze2006}, resulting in a non-neglible photocurrent (see the Supplemental Material section I \cite{Supplmat}).
The excited charges partly screen the internal electric field responsible for the quantum confined stark effect in the QD, leading to the observed blueshift of the QD transitions.

Furthermore, the fact that even for a very low cavity mean photon number of $\sim0.001$ ($P_{in}=11$ pW) nonlinearities take place, indicates that the oxide aperture must form an efficient charge memory compared to the QD-cavity decay rate (77 ns$^{-1}$).
It is well known that the interface between GaAs and aluminum oxide (AlO$_x$), produced by wet oxidation of AlAs, provides a very high density of charge traps, in the form of amorphous oxide and micro crystallites \cite{Twesten1996}, and in the form of elemental interfacial As \cite{Ashby1997}, leading to spatially non-uniform Fermi level pinning \cite{Chen1995}.
The time-resolved charge decay measurements presented below confirm that the charges are relatively long-lived on a $\sim10$ ms timescale.
Also, a comparison with QDs observed outside of the oxide aperture region show that laser induced blueshift is neglible in these regions (see Supplemental Material section II \cite{Supplmat}).
Furthermore, the charging hypothesis agrees with the observed $\propto P^{0.35}$ sublinear power-law, which would be the consequence of Coulomb repulsion between trapped charges, possibly in combination with increased filling of trap states that have an increasing decay rate.

Our results agree with observations in Ref. \cite{Houel2012}, where single--charge fluctuations are probed at a GaAs/AlAs interface located about 50 nm away from the QDs, while a gradual variation was observed for charge traps more than 150 nm away.
We did not observe any single--charge influences as the distance between the QDs and the oxide aperture is about 200 nm in our sample.
In contrast to our work, the GaAs/AlAs interface studied in Ref. \cite{Houel2012} is located in the intrinsic region, while the aperture in our sample is located in the p-doped region, but a similar QD blueshift for increasing laser intensity and constant bias voltage was observed.

\begin{figure}[t]
\centering
\centerline{\includegraphics[angle=0]{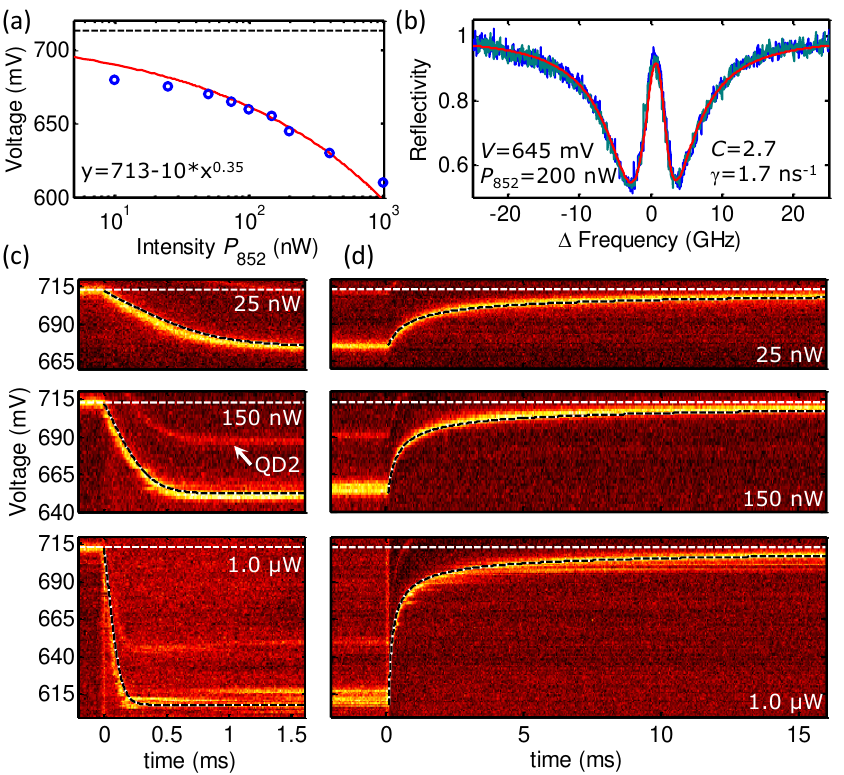}}
\caption{Two-laser scans with an off-resonant ($\lambda = 852$ nm) laser. (a) shows the bias voltage where the QD transition is resonant with the cavity mode, as function of the intensity $P_{852}$ of the off-resonant laser. The black dashed line indicates the QD bias voltage when no off-resonant laser is applied. (b) Example of resonant reflection scans using a weak (1 pW) probe laser, recorded in the presence of an off-resonant laser. Text in the figures denotes the applied bias voltage $V$, intensity $P_{852}$, and QD cooperativity $C$ and dephasing rate $\gamma$ of the predicted red line.
(c,d) Colormaps of the reflected intensity of the probe laser fixed to the center of the cavity resonance (light/dark: high/low signal, corresponding to an on(off) resonant QD), as function of the time and applied bias voltage, for various off-resonant pump intensities. The off-resonant laser is turned on (c) and off (d) at $t = 0$ ms.  The red line in (a) and the black-white dashed in (c,d) are reproduced using the same parameters (see main text for explanation). Note that the time axes of (c,d) are different.}
\label{Fig4}
\end{figure}

Finally, we performed a set of measurements to gain insight in the temporal dynamics of the charge build-up and decay.
For this purpose we use a laser ($\lambda = 852$ nm) that is off-resonant with the cavity but is resonant with the wetting layer.
At a larger laser intensity, more charges are excited compared to the resonant laser and larger QD shifts can be obtained.
We now use the coupled QD-cavity system as a very sensitive probe of the internal electric field, a principle that was also used to monitor a single charge trap in real-time \cite{Arnold2014}.

Figure \ref{Fig4} (a) shows, for various off-resonant pump laser intensities, the bias voltage that has to be applied to tune the QD to the cavity resonance, determined by using a weak ($\sim$1 pW) resonant probe laser.
Again, as in Fig. \ref{Fig2} (d), a clear sub-linear behavior $\propto P_{in}^\beta$ is visible.
At an even higher intensity of 100 $\mu$W, a bias voltage of 350 mV had to be applied to tune the QD in to resonance, clearly indicating that the charge buildup and QD blueshift is not easily saturated.
Strikingly, even for an excitation power of 200 nW, the QD has the same cooperativity $C$ and dephasing rate $\gamma$ as when no off-resonant laser is present, see Fig.~\ref{Fig4}~(b).
This indicates that, even though many charges are excited, they are located relatively far away and give rise to a more-or-less constant effective electric field, thereby preserving the QD coherence.

In order to directly monitor the time dynamics of the charge buildup and decay, we fix a weak probe laser that will not excite any additional charges at the cavity resonance, and monitor the reflectivity as function of bias voltage and time.
Figure \ref{Fig4} (c) shows, for various off-resonant pump laser intensities, reflectivity colormaps of the probe laser when the off-resonant pump laser is turned on at $t = 0$ ms.
For $t<0$ ms, the QD is resonant with the cavity resonance at $V = 713$ mV (white dashed lines), corresponding with a high probe laser reflectivity at this voltage and a low reflectivity at different voltages.
The reflectivity at $V = 713$ mV abruptly decreases as the pump laser is turned on, but is then restored at lower bias voltages, demonstating direct probing of the charge build-up.
For increasing pump laser intensities, the reflectivity is restored at a lower voltage, but also the time for the charge build-up to reach an equilibrium decreases.

The decay of the charges is monitored in Fig. \ref{Fig4} (d), where the pump laser is turned off at $t = 0$ ms.
For $t<0$ ms the QD reflectivity signal is now highest for a low bias voltage when the charge reservoir is saturated.
After the pump is turned off and the charges disappear, the reflectivity now gets restored at an increasing bias voltage.
The charge decay rate is initially fastest, when many charges are still present and the QD is resonant at a lower bias voltage, but then strongly decreases and finally occurs on a $\sim$10 ms timescale, much slower than the charge build-up rate.

We now introduce a simple power-law model to describe the charge build-up and decay dynamics.
We assume that the QD voltage shift $\Delta V$ is proportional to the number of trapped charges $Q$, and increases as a function of the pump laser intensity $P$, such that $\Delta V = -Q= -\Gamma P^{\beta}$.
As a result, the charge decay and buildup is described using: $\frac{dQ}{dt} = P - (Q/\Gamma)^{1/\beta}$.
The red line in Fig. \ref{Fig4} (b) and the white-black dashed lines Fig. \ref{Fig4} (c-d) are reproduced with $\Gamma = 10$ and the same power-law scaling factor $\beta=0.35$ as found earlier, and describe the data nicely.

The effect we found enables a novel method to tune QD transitions without the need for electrical contacts, or enable independent tuning of QDs sharing the same voltage contacts.
Furthermore it could serve as a low power all-optical switch mediated by a charge memory.
The cavity-enhanced feedback mechanism with the charge environment could in principle also occur in other solid-state microcavity structures, where doped layers are present in which charges can be excited and where material defects or interfaces could act as a charge trap.
This interaction therefore also has to be taken into account when studying dynamical nuclear spin polarization (DNP) effects in a cavity, which is of general interest to prolong the QD coherence time and the potential to form a quantum memory.
DNP gives rise to QD line shape modifications, hysteresis and bistabilitity behavior \cite{Braun2006,Tartakovskii2007,Maletinsky2007,Eble2006,Latta2009,Hogele2012, Urbaszek2013}; phenomena that could also give rise to cavity-enhanced feedback mechanisms.


In conclusion, we have studied a neutral QD transition coupled to a microcavity and observed strong cavity-enhanced feedback with the charge environment.
Hysteresis and modifications of the QD line shape are demonstrated at intensities of way less than 1 photon per cavity lifetime, which we explain and model by a QD frequency blueshift attributed to charges trapped at the oxide aperture as a function of the intracavity field.
In general, these results demonstrate the potential of studying and controlling the QD environment using a cavity QED system.

\begin{acknowledgements}
This work was supported by NSF under Grant No. 0960331 and 0901886 and FOM-NWO Grant No. 08QIP6-2.
\end{acknowledgements}
\bibliographystyle{apsrev4-1}

\section{Appendix}
In this Supplemental Material we will first present photocurrent measurements.
To confirm that a $\lambda\sim940$ nm resonant laser excites charges in the cavity region, we compare the photocurrent with when a $\lambda = 852$ nm off-resonant laser is used.
Then we compare photoluminescence data with and without the presence of an aperture, to demonstrate that the oxide aperture plays an important role in trapping the charges,.
These observations support our claim that the mechanism behind the observed nonlinear effects in the resonant QD scans, is the excitation of charges that are trapped by the oxide aperture.

\subsection{I. Photocurrent measurements}
\begin{figure}[h]
\centering
\centerline{\includegraphics[angle=0,scale = 1.0]{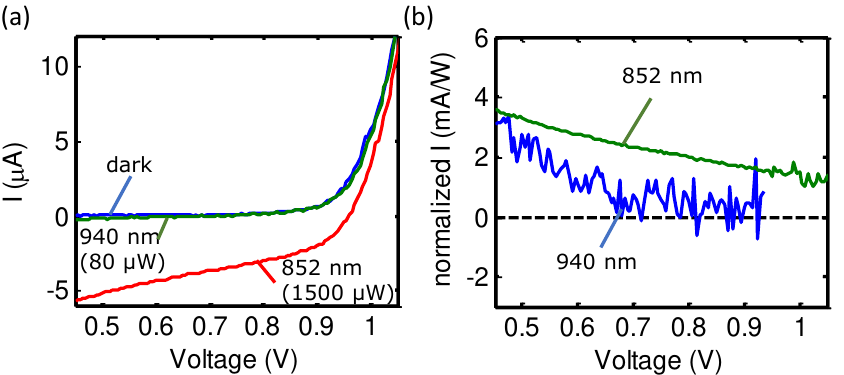}}
\caption{(a) The current through the sample as function of the applied bias voltage without (dark) and with the presence of a 852 nm or $\sim940$
nm laser. The 940 nm laser was locked at the microcavity resonance. The used intensity is mentioned between brackets.(b) Difference between the current when no laser and when an 852 nm or 940 nm laser is present, normalized by the input laser intensity.}
\label{SupplFig1}
\end{figure}

Current as function of bias voltage was measured without (dark) and with the presence of a $\lambda=852$ nm or a $\lambda\sim940$ nm laser, see Fig. \ref{SupplFig1} (a).
The $\lambda\sim940$ nm laser frequency was locked at the cavity resonance, such that the absorption is increased due to the high finesse ($\sim2\times10^3$) of the cavity, as is the case during the resonant scans in the main text.
When no additional pump laser is applied, a current flows through the device in the same direction as that the forward bias voltage is applied, giving rise to an IV-curve that is typical for a PIN diode device.
In the presence of an additional pump laser charges are now optically excited and flow in the opposite direction due to the internal field being present, giving rise to an additional negative photocurrent.
In Fig. \ref{SupplFig1} (b) we compare the absolute value of the photocurrent produced by the $\lambda \sim 940$ nm and $\lambda = 852$ nm laser, normalized by the pump laser intensity.
For both lasers a strong bias voltage dependency is visible and the collected photocurrent is smaller at a larger bias voltage, which is closer to the flatband regime.
This indicates that either the absorption, or the collection efficiency of the excited charges, is voltage dependent.
The 940 nm laser gives rise to a photocurrent of about 0.5 mA/W at a bias voltage of 700 mV where charge neutral QDs are typically operated, which at an intensity of $1$ nW corresponds to about $3\times10^6$ charges being excited and collected per second.

\subsection{II. Photoluminescence measurements}
\begin{figure}[h]
\centering
\centerline{\includegraphics[angle=0,scale = 1.0]{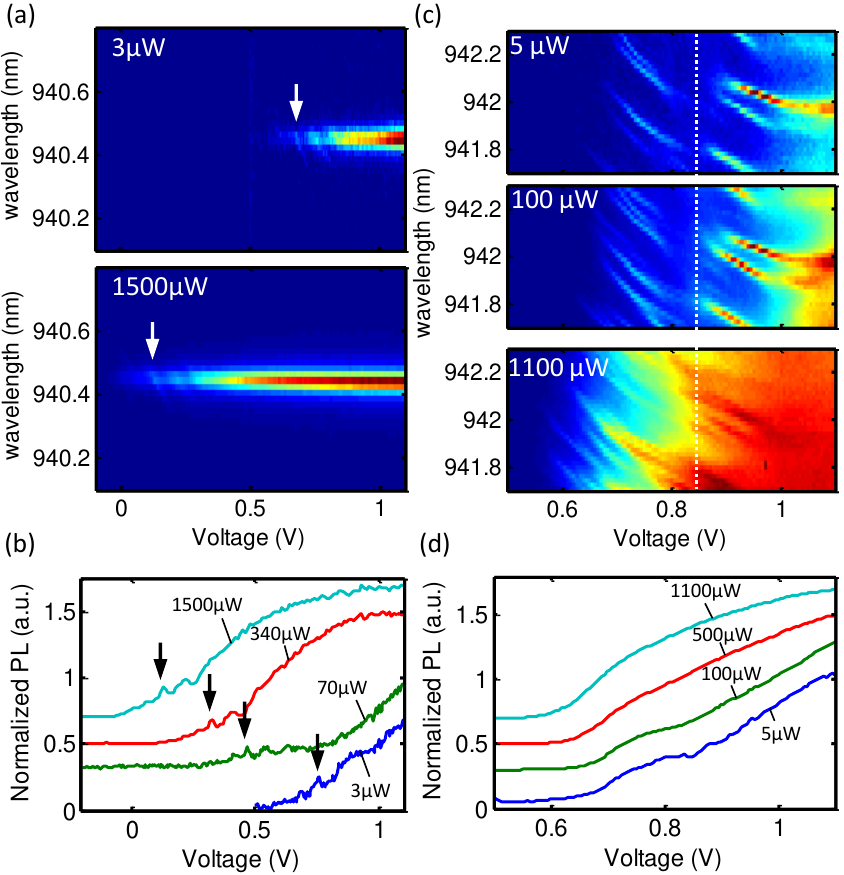}}
\caption{Photoluminescence (PL) colormaps as function of the applied bias voltage and collection wavelength recorded at (a) the microcavity, and (c) from an unbalanced planar cavity. The intensity of the ($\lambda = 852$ nm) pump laser is indicated in the top left corner of the figures. (b) PL intensity as function of bias voltage at the cavity resonance, for various pump laser intensities. The vertical arrows in (a, b) denotes the same QD peak. (d) average PL intensity as function of voltage collected from the unbalanced planar cavity. The white dashed vertical line serves as a reference. An offset has been added between the curves in (b, d).}
\label{SupplFig2}
\end{figure}

Next, we compare the photoluminescence collected from the oxide-aperture microcavity and from a region with an unbalanced cavity.
In the unbalanced cavity region the top DBR has been nearly completely etched away, in order to enhance the collection efficiency, and no oxide aperture has been applied.
Figure \ref{SupplFig2} (a,c) present colormaps of the PL spectra as a function of applied bias voltage and collected wavelength, using a $\lambda = 852$ nm pump laser of various intensities.
Figure \ref{SupplFig2} (b) shows the PL intensity as function of voltage at the cavity resonance.
Single QD lines (marked by the arrows) are hardly distinguishable from the background cavity emission, which originates from broadband wetting layer emission. 
For increasing pump laser intensity, it is clearly visible that the QD lines appear at a lower bias voltage.
For the QDs in the region without an oxide aperture, the shift in voltage is however much smaller.
The QDs in Fig.~\ref{SupplFig2}~(c) are now better visible, even though at high pump laser intensities for increasing voltage again broadband emission 'washes out' clear QD signatures.
Figure \ref{SupplFig2} (d) presents the mean PL counts for various pump intensities.
A shoulder is visible, which corresponds with the voltage range where the QDs are on average charge neutral.
The shoulder does shift to a lower voltage range, but the shift (in total $\sim50$ mV for 1100 $\mu$W) is much smaller compared to the total $\sim 700$ mV (for 1500 $\mu$W) voltage shift of the QDs in the microcavity.
Even though it is likely that roughly the same amount of charges are excited by the $\lambda=852$ nm laser in the microcavity region compared to elsewhere in the sample, these results indicate that the oxide aperture provides a charge memory, causing a much larger effective charge buildup resulting in a larger voltage shift of the QDs.

\end{document}